\documentclass[iop]{emulateapj}
\usepackage{amsmath}
\usepackage{times}
\usepackage{graphics, subfigure}
\usepackage{epsfig}
\usepackage{multirow}
\usepackage{ulem}
\usepackage{pdfpages}

\def\simlt{\mathrel{\hbox{\rlap{\hbox{\lower4pt\hbox{$\sim$}}}\hbox{$<$}}}}
\def\simgt{\mathrel{\hbox{\rlap{\hbox{\lower4pt\hbox{$\sim$}}}\hbox{$>$}}}}
\newcommand{\Mshell}{M_{\textrm{shell}}}
\newcommand{\Mshellmax}{M_{\textrm{shell},\textrm{max}}}
\newcommand{\frot}{f_{\textrm{rot}}}
\newcommand{\frotz}{f_{\textrm{rot},0}}
\newcommand{\rrotz}{r_{\textrm{rot},0}}
\newcommand{\xrotz}{x_{\textrm{rot},0}}

\begin{document}

\title{Thermonuclear explosion of rotating massive stars could explain core-collapse supernovae}

\author{Doron Kushnir\altaffilmark{1}} \altaffiltext{1}{Institute for Advanced Study, Einstein Drive, Princeton, NJ, 08540, USA, kushnir@ias.edu}

\begin{abstract}
It is widely thought that core-collapse supernovae (CCSNe), the explosions of massive stars following the collapse of the stars' iron cores, is obtained due to energy deposition by neutrinos. So far, this scenario was not demonstrated from first principles.
Kushnir and Katz (2014) have recently shown, by using one-dimensional simulations, that if the neutrinos failed to explode the star, a thermonuclear explosion of the outer shells is possible for some (tuned) initial profiles. However, the energy released was small and negligible amounts of ejected $^{56}$Ni were obtained, implying that these one-dimensional collapse induced thermonuclear explosions (CITE) are unlikely to represent typical CCSNe. Here I provide evidence supporting a scenario in which the majority of CCSNe are the result of CITE. I use two-dimensional simulations to show that collapse of stars that include slowly (few percent of breakup) rotating $\sim0.1-10\,M_{\odot}$ shells of mixed helium-oxygen, leads to an ignition of a thermonuclear detonation wave that unbinds the stars' outer layers. Simulations of massive stars with different properties show that CITE is a robust process, and results in explosions with kinetic energies in the range of $10^{49}-10^{52}\,\textrm{erg}$, and $^{56}$Ni yields of up to $\sim\,M_{\odot}$, which are correlated, in agreement with observations for the majority of CCSNe. Stronger explosions are predicted from higher mass progenitors that leave more massive remnants, in contrast to the neutrino mechanism. Neutron stars are produced in weak ($\simlt10^{51}\,\textrm{erg}$) explosions, while strong ($\simgt10^{51}\,\textrm{erg}$) explosions leave black hole remnants.
 \end{abstract}



\section{Introduction}
\label{sec:Introduction}

There is strong evidence that CCSNe are explosions of massive stars, involving the collapse of the stars' iron cores \citep[][]{BBFH,87a,smartt2009} and ejection of the outer layers. It is widely thought that the $\sim10^{51}\,\textrm{erg}$ observed kinetic energy (KE) is obtained due to the deposition of a small fraction ($\sim1\%$) of the gravitational energy ($\sim10^{53}\,\textrm{erg}$) released in neutrinos \citep[see][for reviews]{Bethe90,Burrows2013}. So far, this scenario was not demonstrated from first principles.

\citet[][]{BBFH} suggested a different mechanism for the explosion during core-collapse that does not involve the emitted neutrinos. In the proposed scenario, increased burning rates due to adiabatic heating of the outer shells as they collapse lead to a thermonuclear explosion \citep[see also][]{Hoyle60,Fowler64}. This collapse-induced thermonuclear explosion (CITE) naturally produces $\sim10^{51}\,\textrm{erg}$ from thermonuclear burning of $\sim M_{\odot}$ (gain of $\sim \textrm{MeV}/m_{p}$). A few one-dimensional studies suggested that CITE fails \citep[][]{Colgate66,WW82,Bodenheimer}. It was recently shown, by using one-dimensional simulations, that CITE is actually possible for some (tuned) initial density and composition profiles \citep[][]{Kushnir14}, which include shells of mixed helium and oxygen. An ignition of a thermonuclear detonation is obtained, $\sim10$ seconds after the core collapsed, that unbinds the outer layers of the star. However, the energy released was small, $\lesssim10^{50}\,\textrm{erg}$, and negligible amounts of ejected $^{56}$Ni were obtained, implying that these one-dimensional CITE are not typical CCSNe.

Here I provide evidence supporting a scenario in which the majority of CCSNe are CITE of rotating massive stars. By using two-dimensional simulations with a fully resolved ignition process, I show that collapse of stars that include slowly (few percent of breakup) rotating $\sim0.1-10\,M_{\odot}$ shells of He-O with densities of $\textrm{few}\times10^{3}\,\textrm{g}\,\textrm{cm}^{-3}$, leads to an ignition of a thermonuclear detonation that unbinds the stars' outer layers.

Previous studies of thermonuclear explosions of collapsing rotating stars \citep[][]{Bodenheimer,collapsar}, without He-O mixtures, did not result in an ignition of a detonation. I show that the presence of He-O mixtures is a necessary condition for explosion, making the previous studies consistent with the present work.

Section~\ref{sec:numerical} describes the simulations. Section~\ref{sec:observations} demonstrates the consistency of CITE with the majority of CCSNe, including new predictions.

\section{Numerical simulations}
\label{sec:numerical}

Section~\ref{sec:initial one-dimensional} describes the initial profiles, Section~\ref{sec:simulation} describes the numerical tool, Section~\ref{sec:example} presents a simulation of a successful thermonuclear explosion that leads to a typical CCSN, and Section~\ref{sec:sensitivity} demonstrates that CITE is robust, by examining the sensitivity of the results to the assumed initial profile.

\subsection{Initial profiles}
\label{sec:initial one-dimensional}

Pre-collapse profiles are uncertain \citep[see, e.g][]{Smith2014}, but nevertheless stellar evolution calculations of rotating massive stars generally predict the existence of a He-O shell \citep[][]{Hirschi2004,Hirschi2005,Hirschi2007,Yusof2013}. Instead of relying on some specific calculated profiles, I systematically examine the explosion's properties as a function of the He-O shell properties.
Profiles are composed as follows (this parametrization is somewhat different from \citet[][]{Kushnir14}):
\begin{itemize}
\item The profile is composed of shells with constant entropy (per unit mass), $s$, constant composition, and in hydrostatic equilibrium.
\item $1.2\,M_{\odot}$ is placed within $r<2\cdot10^{8}\,\textrm{cm}$, representing a degenerate iron core.
\item The entropy of the He-O shell is $s_{\textrm{He-O}}$ and the mass fractions are $X_{\textrm{He}}=1-X_{\textrm{O}}$. At the base of the shell the enclosed mass is $M_{\textrm{base}}$, the density is $\rho_{\textrm{base}}$, and the ratio between the local burning time and the free-fall time, $t_{b}/t_{ff}$, is $t_{b,0}/t_{ff,0}$, where $t_{b}=\varepsilon/\dot Q$, $\varepsilon$ is the internal energy, $\dot Q$ is the thermonuclear energy production rate, and $t_{ff}=r^{3/2}/\sqrt{2GM(r)}$. The inner radius of the shell, $r_{\textrm{base}}$, is set by these parameters, and the outer radius is placed at $\rho=10^{3}\,\textrm{g}\,\textrm{cm}^{-3}$.
\item Pure oxygen (helium) is placed below (above) the He-O shell, where the entropies are determined by requiring enclosed mass of $1.2\,M_{\odot}$ within $r<2\cdot10^{8}\,\textrm{cm}$ and zero pressure or low temperature ($10^5\,\textrm{K}$) at the profile's fixed outer radius of $3\cdot 10^{10}\,\textrm{cm}$. 
\end{itemize}
Instead of specifying $s_{\textrm{He-O}}$, it is more convenient to specify $\Mshell$, the mass of the He-O shell. For given $M_{\textrm{base}}$, $\rho_{\textrm{base}}$, and $t_{b,0}/t_{ff,0}$, there is a maximal possible $\Mshell$ (obtained for a minimal possible $s_{\textrm{He-O}}$), denoted by $\Mshellmax$.

Following previous studies \citep[][]{Bodenheimer,collapsar}, the angular momentum is initially distributed such that $\frot$, the ratio of the centrifugal force to the component of gravitational force perpendicular to the rotation axis, is some constant $\frotz$ ($0.05$ for most simulations) throughout the profile, except for the following:
\begin{itemize}
\item $\frot=0$ at small radii, $r<\rrotz\equiv\xrotz r_{\textrm{base}}$. $\xrotz=0.8$ for most simulations.
\item $\frot=0$ at large radii, and increases linearly with decreasing radius between $r=2\cdot10^{10}\,\textrm{cm}$ and $r=1\cdot10^{10}\,\textrm{cm}$ to $\frotz$. This is done for numerical stability, and has a small effect on the results.
\end{itemize}
The initial profiles are chosen to be in hydrostatic equilibrium without rotation. Rotation is added such that initially the profile is slightly out of equilibrium (see Section~\ref{sec:frotz}). 

\subsection{Collapse Simulations}
\label{sec:simulation}

The problem considered is axisymmetric, allowing the use of two-dimensional numerical simulations with high resolution. I employ the FLASH4.0 code with thermonuclear burning \citep[Eulerian, adaptive mesh refinement;][]{dubey2009flash} using cylindrical coordinates $(R,z)$ to calculate one quadrant, with a resolution (i.e. the minimal allowed cell size within the most resolved regions) of $\sim\textrm{few}\times10\,\textrm{km}$. I use the 13 isotope alpha-chain reaction network of VULCAN2D \citep[][]{Livne1993IMT}, and a burning-limiter \citep[][]{Kushnir2013}.

Angular momentum was implemented similarly to previous studies \citep[][]{Lindner2010,Fernandez2013}. Specific angular momentum is included as a mass scalar quantity. The centrifugal force is included as part of the default treatment of ``fictitious'' forces that arise in curvilinear coordinates. The specific KE associated with rotation is added to the specific KE used in the regular scheme of FLASH4.0, as necessary when viscous forces are neglected.

I assume that neutrinos escape with negligible effect on the outer layers (gravitational mass loss is only considered for the post-processing estimate of the remnant mass). Layers below the inner boundary, $r_{\textrm{inner}}$ ($5\cdot10^{7}\,\textrm{cm}$ for most simulations), are assumed to have already collapsed, and the pressure within this radius is held at zero throughout the simulation. Mass of material that (freely) flows through the boundary is added to the original collapsed mass and contributes to the gravitational field.

At locations in the progenitor where $T>2\cdot 10^9\,\textrm{K}$, oxygen is replaced with silicon to prevent fast initial burning, and the region between $r_{\textrm{inner}}$ and $r=2\cdot10^{8}\,\textrm{cm}$ is filled with $s=1\,k_{B}$  iron (satisfying hydrostatic equilibrium), which is prevented from burning. These regions collapse rapidly thorough the inner boundary and have negligible effect on the results.

For simulations in which an explosion was obtained, the original resolution was reduced successively by factors of $2$ for each doubling of the shock radius, starting with the first resolution increase made at a chosen threshold radius  $r_{\textrm{res}}$ ($5\cdot10^{9}\,\textrm{cm}$ for most simulations).

\subsection{CITE is possible - example of a successful explosion which leads to a typical CCSN}
\label{sec:example}

\begin{figure}
\includegraphics[width=0.5\textwidth]{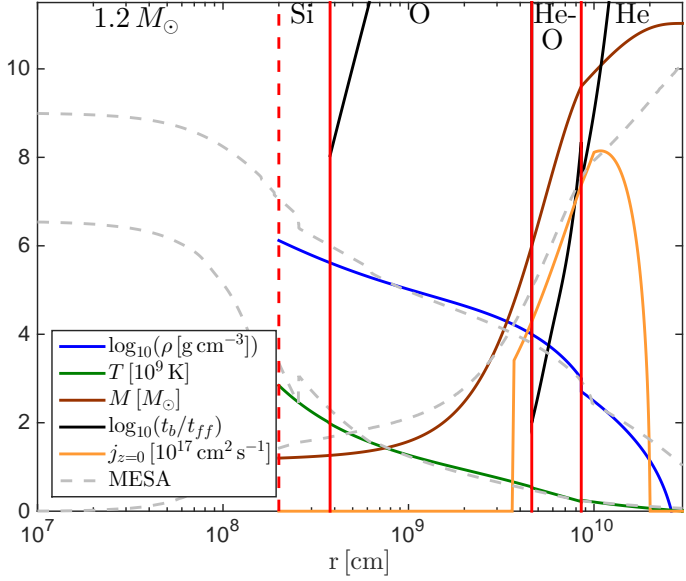}
\caption{A pre-collapse profile (density, temperature, enclosed mass, burning to free-fall time ratio, and specific angular momentum on the equatorial plane ($j_{z=0}$)) that leads to a typical CCSN. For this profile $M_{\textrm{base}}=6\,M_{\odot}$, $\rho_{\textrm{base}}=10^{4}\,\textrm{g}\,\textrm{cm}^{-3}$, $t_{b,0}/t_{ff,0}=100$, $X_{\textrm{He}}=X_{\textrm{O}}=0.5$ and $\Mshell=\Mshellmax\approx3.60\,M_{\odot}$, leading to $r_{\textrm{base}}\approx4.62\cdot10^{9}\,\textrm{cm}$, burning time at $r_{\textrm{base}}$ of $t_{b,0}\approx7.9\cdot10^{2}\,\textrm{s}$ and a total mass of $\approx11.0\,M_{\odot}$ (the profile below $r=2\cdot10^{8}\,\textrm{cm}$ has negligible effect on the results and is not shown). The obtained density, temperature, and enclosed mass profiles are similar to pre-collapse profiles of a $30\,M_{\odot}$ star (dashed gray), calculated by Roni Waldman with the MESA stellar evolution code \citep[][]{MESA}. The main difference is the existence of the He-O mixture. The rotation parameters are $\frotz=0.05$ and $\xrotz=0.8$ ($\rrotz=\xrotz\times r_{\textrm{base}}\approx3.69\cdot10^{9}\,\textrm{cm}$).
\label{fig:InitialProfile}}
\end{figure}

\begin{figure}
        \subfigure[]{
             \includegraphics[width=0.45\textwidth]{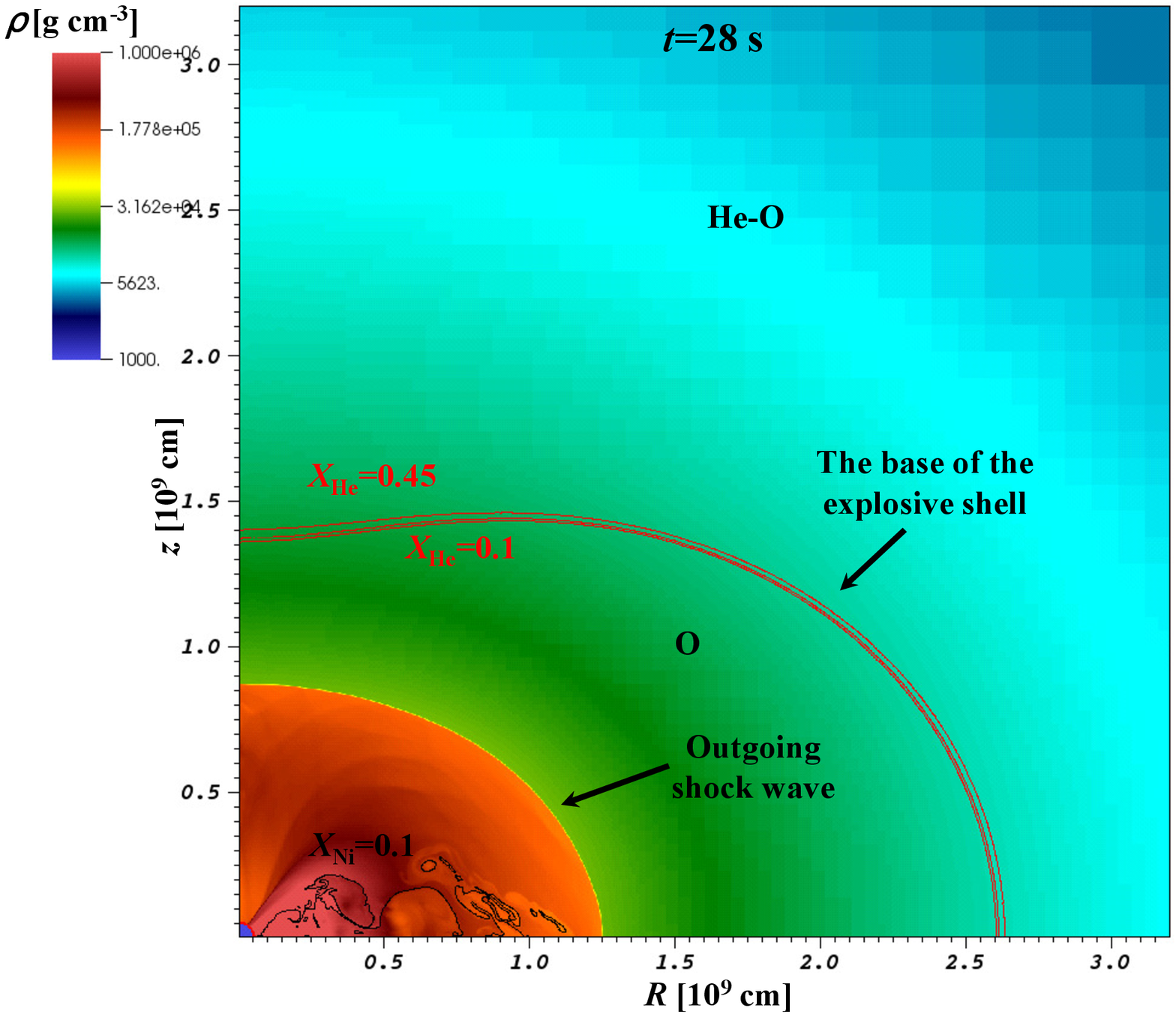}
        }
        \subfigure[]{
             \includegraphics[width=0.45\textwidth]{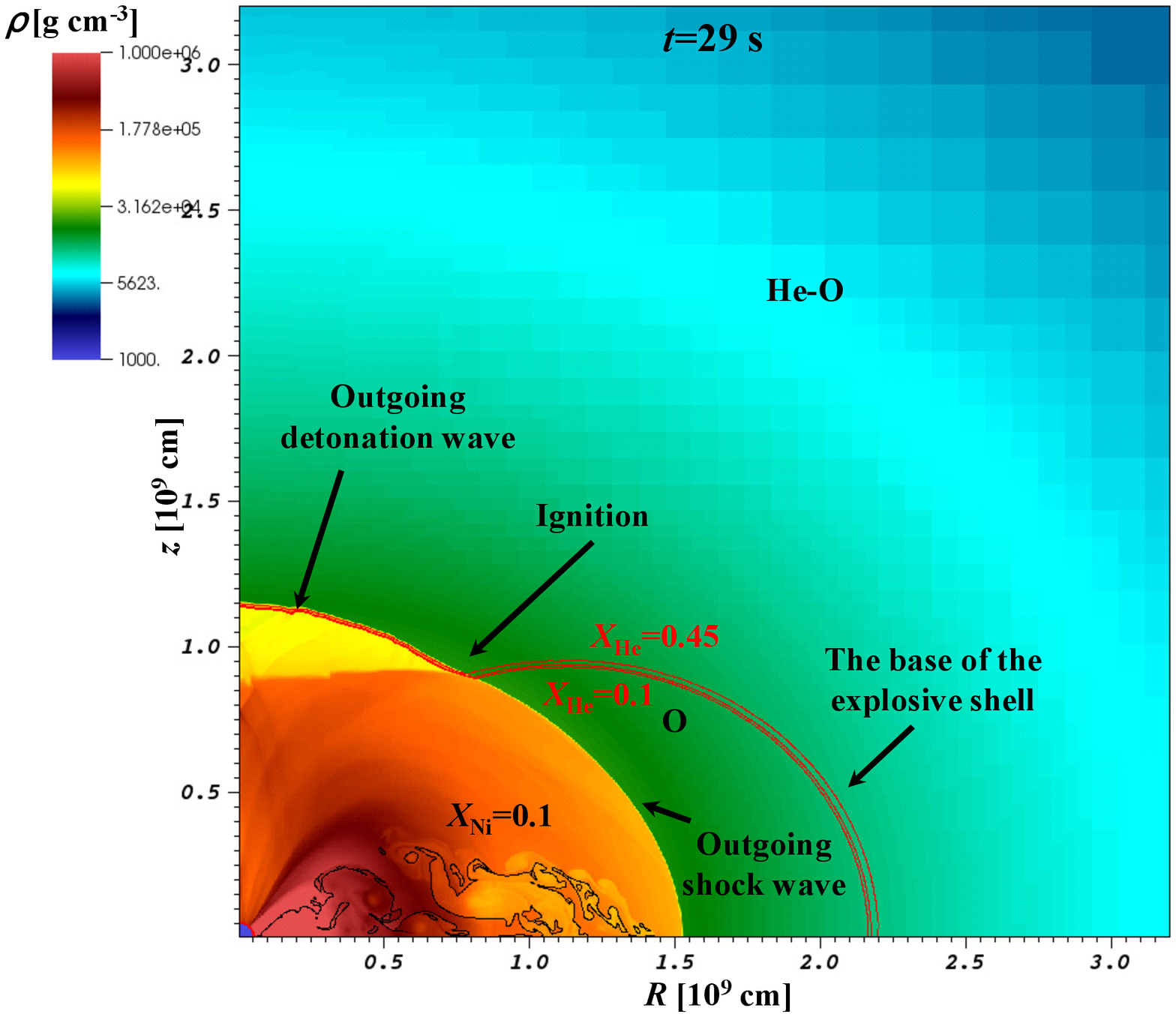}
        }
        \subfigure[]{
             \includegraphics[width=0.45\textwidth]{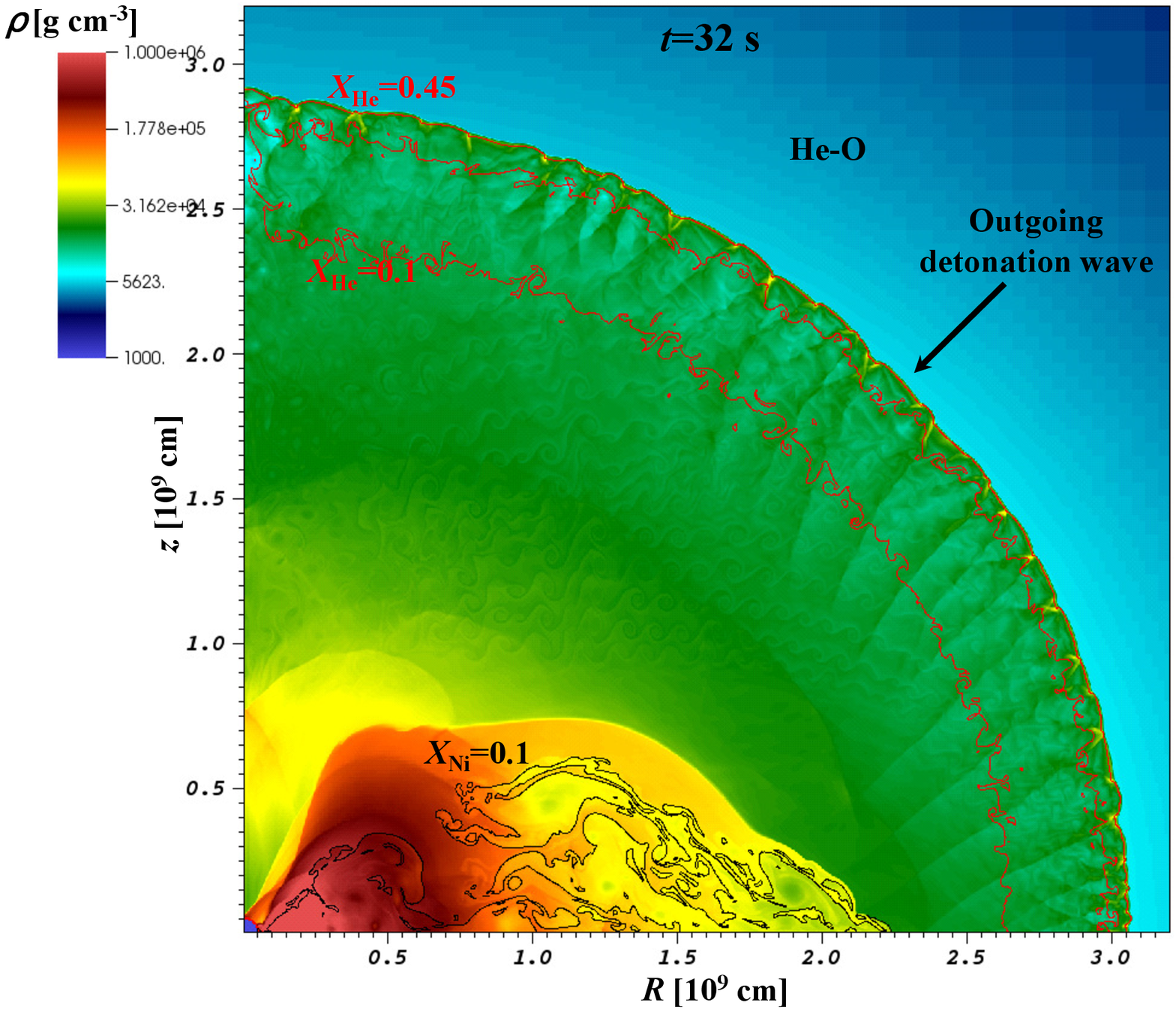}
        }
        \subfigure[]{
             \includegraphics[width=0.45\textwidth]{f2d.eps}
        }
      \caption{Dynamical evolution of the collapse for the initial conditions of Figure~\ref{fig:InitialProfile}. Panels (a-c): logarithmic density maps at different times since collapse with isotope contours of He (red, $X_{\textrm{He}}=0.1,0.2,0.45$) and $^{56}$Ni (black, $X_{\textrm{Ni}}=0.1$). Panel (d): rate of thermonuclear energy production, $\dot{E}_{\textrm{burn}}$ (red), accumulated thermonuclear energy produced, $E_{\textrm{burn}}$ (blue), and total KE of mass elements with positive radial (spherical) velocity, $E_{\textrm{kin}}(v_{r}>0)$ (black).}
   \label{fig:Evolution}
\end{figure}

A pre-collapse profile which leads to a typical CCSN is shown in Figure~\ref{fig:InitialProfile}. The dynamical evolution of the collapse, calculated with a resolution of $\approx30\,\textrm{km}$, is shown in Figure~\ref{fig:Evolution}. Collapsing material that includes angular momentum reaches a centrifugal barrier at a radius smaller by $\simeq \frotz$ of its initial radius, leading to an increased pressure and to the formation of a rotation-induced accretion shock (RIAS), seen in panel (a). The in-fall velocity of the shock-heated oxygen is significantly reduced, allowing synthesis of $^{56}$Ni (black contour) within the hot and dense downstream conditions. The energy released in this thermonuclear processing ($\simeq3\cdot10^{50}\,\textrm{erg}$ until $t=28\,\textrm{s}$, panel (d)) is sufficient for an expansion of the synthesized material. The base of the He-O shell, seen in panel (a), is deformed during collapse from its initial spherical shape. The RIAS hits the He-O shell at $t\simeq29\,\textrm{s}$ which causes an ignition of a detonation (panel (b)). Because the shapes of the RIAS and the base of the He-O shell are different, the ignition position propagates from the symmetry axis towards the equator. The detonation  propagates outward (panel (c)), producing thermonuclear energy at $\textrm{few}\times 10^{50}\,\textrm{erg}\,\textrm{s}^{-1}$ (panel (d)). The pressure built from accumulating thermonuclear energy, aided by the RIAS that acts as a piston, manages to halt the inward collapse and causes an expansion that leads to an outward motion. Once the detonation reaches outer layers with densities $\rho\lesssim\textrm{few}\times10^{3}\,\textrm{g}\,\textrm{cm}^{-3}$ it decays and transitions to a hydrodynamic shock which continues to propagate outwards. 

The resulting ejecta has a mass of $\approx4.5\,M_{\odot}$ (leaving $\approx6.5\,M_{\odot}$ of material with negative spherical radial velocity, hereafter \textit{in-falling mass}), $\textrm{KE}\approx1.3\cdot10^{51}\,\textrm{erg}$, and a $^{56}$Ni mass, $M_{\textrm{Ni}}$, of $\simeq0.08\,M_{\odot}$. The obtained KE and $M_{\textrm{Ni}}$ are typically observed in CCSNe (see Section~\ref{sec:observations}). The KE of the ejecta may change slightly if a hydrogen envelope is added.

\subsubsection{Numerical convergence}
\label{sec:convergence}

\begin{figure}
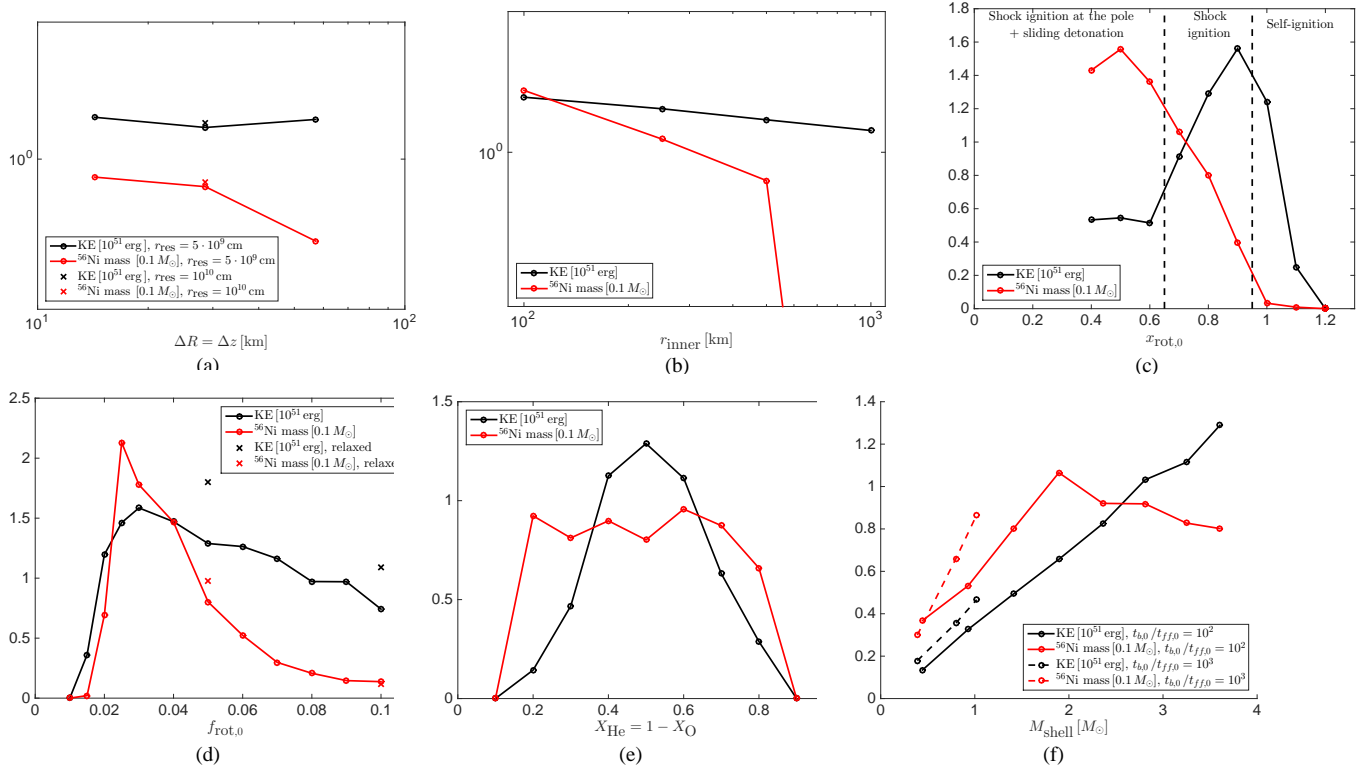

        \subfigure[]{
             \includegraphics[width=0.3\textwidth]{f3a.eps}
        }
        \subfigure[]{
             \includegraphics[width=0.3\textwidth]{f3b.eps}
        }
        \subfigure[]{
             \includegraphics[width=0.3\textwidth]{f3c.eps}
        }
        \subfigure[]{
             \includegraphics[width=0.3\textwidth]{f3d.eps}
        }
        \subfigure[]{
             \includegraphics[width=0.3\textwidth]{f3e.eps}
        }
        \subfigure[]{
             \includegraphics[width=0.3\textwidth]{f3f.eps}
        }
      \caption{KE (black) and $M_{\textrm{Ni}}$ (red) as a function of several numerical and physical parameters. panel (a): resolution, x-symbols are obtained from simulation with $r_{\textrm{res}}=10^{10}\,\textrm{cm}$; panel (b): $r_{\textrm{inner}}$; panel (c): $\xrotz$; panel (d): $\frotz$, x-symbols are obtained from simulation with relaxed initial models; panel (e): composition; panel (f): $\Mshell$, for $t_{b,0}/t_{ff,0}=10^{2}$ (solid) and $t_{b,0}/t_{ff,0}=10^{3}$ (dashed).}
   \label{fig:example_param}
\end{figure}

The KE and $M_{\textrm{Ni}}$ as a function of resolution are shown in panel (a) of Figure~\ref{fig:example_param}. The results are converged to $\approx10\%$ for resolution of $\approx30\,\textrm{km}$, used throughout the paper.

The $^{56}$Ni yield is estimated as the total mass of $^{56}$Ni with positive (spherical) radial velocity, $M_{\textrm{Ni}}(v_{r}>0)$, at the latest time with the original resolution. At this time $M_{\textrm{Ni}}(v_{r}>0)$ is roughly constant with time, and a decrease in $M_{\textrm{Ni}}(v_{r}>0)$ was obtained as the resolution was reduced. In order to verify that the decrease in $M_{^\textrm{Ni}}(v_{r}>0)$ is numerical, another calculation with $r_{\textrm{res}}=10^{10}\,\textrm{cm}$ was performed (x-symbols in panel (a) of Figure~\ref{fig:example_param}), in which similar $M_{\textrm{Ni}}$ was obtained. The reduction of the resolution inhibits accurate calculation of the asymptotic $^{56}$Ni distribution in the ejecta.

The KE and $M_{\textrm{Ni}}$ as a function of $r_{\textrm{inner}}$ are shown in panel (b) of Figure~\ref{fig:example_param}. The smaller $r_{\textrm{inner}}$, the earlier the RIAS is launched. As can be seen the KE and $M_{\textrm{Ni}}$ are not sensitive to the exact value of $r_{\textrm{inner}}$ with the $M_{\textrm{Ni}}$ slowly increasing for smaller $r_{\textrm{inner}}$. It would be desirable to decrease $r_{\textrm{inner}}$ below the minimal value tried here ($100\,\textrm{km}$), however the in-fall velocities at this radius are already $\approx0.4c$, leading to tens of percent error for a non-relativistic calculation. In what follows I employ $r_{\textrm{inner}}=500\,\textrm{km}$, keeping in mind that $M_{\textrm{Ni}}$ is probably under-predicted by a factor of a few. 

\subsection{CITE is robust - the full set of simulations}
\label{sec:sensitivity}

In Sections~\ref{sec:xrotz}--\ref{sec:shell mass} I examine the sensitivity of the results of Section~\ref{sec:example} to $\xrotz$, $\frotz$, $X_{\textrm{He}}$, $\Mshell$, and $t_{b,0}/t_{ff,0}$, while in Section~\ref{sec:shell location} $M_{\textrm{base}}$ and $\rho_{\textrm{base}}$ are varied. The relevant figure for each section is indicated in the title. A list of the simulations is given in 
Tables~\ref{tbl:lista}-\ref{tbl:listb}.

\subsubsection{Sensitivity to the inner boundary of the rotation zone ($\xrotz$) -- panel (c) of Figure~\ref{fig:example_param}}
\label{sec:xrotz}

For $0.6<\xrotz<1$ ignition occurs immediately after the interaction with the RIAS, similarly to panel (b) of Figure~\ref{fig:Evolution}. The smaller $\xrotz$, the earlier the RIAS is launched, leading to smaller KE (since the He-O shell is less compressed) but to higher $M_{\textrm{Ni}}$ (because of the higher density of the oxygen).

For $\xrotz=1$ there is no RIAS inner to the He-O shell, and it ignites because of adiabatic compression during free-fall (\textit{self-ignition}), similarly to one-dimensional simulations \citep[][]{Kushnir14}. In fact, in a one-dimensional simulation of this profile, while the explosion fails, $\approx2.5\cdot10^{51}\,\textrm{erg}$ of thermonuclear energy is released, which is comparable to the $\xrotz=1$ case ($\approx2.9\cdot10^{51}\,\textrm{erg}$). The one-dimensional case fails because only a small fraction of the thermonuclear energy is converted to outward motion and cannot overcome the binding energy of the star \citep[][]{Kushnir14}, $E_{\textrm{bin}}\approx-9\cdot10^{50}\,\textrm{erg}$ (exterior to the base of the He-O shell, corrected for thermal energy). For $\xrotz=1$, the RIAS formes inside the He-O shell after ignition, and reaches the detonation front while acting as a piston that increases the fraction of thermonuclear energy that is converted to outward motion ($\approx0.75$ in this case, $\simlt0.35$ in the equivalent one-dimensional case). Small amounts of $^{56}$Ni are synthesized, since the RIAS forms late, where only a small amount of high density material is present. As $\xrotz$ is increased, a smaller fraction of the He-O shell is RIAS supported, and for $\xrotz=1.2$ the explosion fails, similarly to the one-dimensional case. 

For $\xrotz<0.7$ more complicated dynamics are obtained, since the RIAS hits the He-O shell at early times, in which the density is too low for immediate ignition. However, later on ignition happens at the pole, where the induction time is shortest, and then the detonation slides towards the equator. The KE is smaller by a factor of a few compared to shock ignition, but the $M_{\textrm{Ni}}$ is somewhat higher because of the high density material that is being shocked.

\subsubsection{Sensitivity to the rotation speed ($\frotz$) -- panel (d) of Figure~\ref{fig:example_param}}
\label{sec:frotz}

The KE depends weakly on $\frotz$, as long as $\frotz\ge0.02$. For smaller values, KE drops sharply, and no explosion is obtained for $\frotz=0.01$, as no RIAS was launched at relevant times. The $M_{\textrm{Ni}}$ decreases with increasing $\frotz$, since the RIAS is launched at larger radii, leading to collapsing material with smaller density. This demonstrates that rotation is required for CITE.

To test the effect of the artificial departure from hydrostatic equilibrium due to rotation, I preformed a set of simulations where the initial profiles were first relaxed to equilibrium before the collapse. The relaxation was run for $\sim1000\,\textrm{s}$ without inflow through $r_{\textrm{inner}}$ and without burning. The results of such simulations ($\frotz=0.05$ and $\frotz=0.1$) change by $\simlt50\%$ (x-symbols).

\subsubsection{$X_{\textrm{He}}$ -- panel (e) of Figure~\ref{fig:example_param}}
\label{sec:composition}

For low (high) values of $X_{\textrm{He}}$ the KE decreases since the energy content of the He-O mixture is decreasing (since there are not enough target nuclei for alpha capture and the energy content of the He-O mixture cannot be extracted efficiently), until the explosions fails for $X_{\textrm{He}}=0.1(0.9)$. This demonstrates the He-O mixture requirement for CITE. $^{56}$Ni is produced below the He-O shell, and is not affected by its composition.

\subsubsection{$\Mshell$ and $t_{b,0}/t_{ff,0}$ -- panel (f) of Figure~\ref{fig:example_param}}
\label{sec:shell mass}

Decreasing $\Mshell$ decreases both the available thermonuclear energy and the binding energy $|E_{\textrm{bin}}|$. However, the obtained KE roughly follows $|E_{\textrm{bin}}|$ (see Section~\ref{sec:shell location}). $M_{\textrm{Ni}}$ is not sensitive to $\Mshell$, unless weak explosion is obtained. For $t_{b,0}/t_{ff,0}=10^{3}$ the maximal He-O shell mass is only $\Mshellmax\approx1.02\,M_{\odot}$, but at a given $\Mshell$ the results depend weakly on $t_{b,0}/t_{ff,0}$.

\subsubsection{$M_{\textrm{base}}$ and $\rho_{\textrm{base}}$ - Figure~\ref{fig:full}}
\label{sec:shell location}

In this section $t_{b,0}/t_{ff,0}=100$, $X_{\textrm{He}}=0.5$, $\Mshell=\Mshellmax$, $\frotz=0.05$, $\xrotz=0.8$, and the ranges $M_{\textrm{base}}\in\left[1.5,16.5\right]M_{\odot}$, $\rho_{\textrm{base}}\in\left[0.5,2.5\right]\cdot10^{4}\,\textrm{g}\,\textrm{cm}^{-3}$ are scanned.

The KE, $M_{\textrm{Ni}}$, and in-falling mass are shown in Figure~\ref{fig:full} (only successful explosions are presented). The maximal possible KE ($M_{\textrm{Ni}}$) increases with $M_{\textrm{base}}$, and ranges between $10^{49}\,\textrm{erg}$ to $10^{52}\,\textrm{erg}$ (negligible amounts to $\approx0.5\,M_{\odot}$), covering the observed range for the vast majority of CCSNe (see Section~\ref{sec:observations}). The in-falling mass is larger by $0.1-15\,M_{\odot}$ than $M_{\textrm{base}}$.

As can be seen in panel (d) of Figure~\ref{fig:full}, the KE never exceeds the binding energy $|E_{\textrm{bin}}|$ by more than a factor of $2.5$. This can be understood by comparing the available thermonuclear energy $\sim\Mshell\times\textrm{MeV}/m_{p}$ with the binding energy $\sim -G M_{\textrm{base}}\Mshell/r_{\textrm{base}}$, and noting that $\textrm{few}\times G M_{\textrm{base}}/r_{\textrm{base}}\approx\textrm{MeV}/m_{p}$ \citep[][]{Kushnir14}. In the absence of tuning between the released thermonuclear energy and $E_{\textrm{bin}}$, the minimal KE cannot be much smaller than $|E_{\textrm{bin}}|$ ($\textrm{KE}\simgt0.25|E_{\textrm{bin}}|$ for all calculations). Therefore, $\textrm{KE}\sim|E_{\textrm{bin}}|$ for CITE, in contrast to the neutrino mechanism, where larger KE are obtained for smaller $|E_{\textrm{bin}}|$ \cite[][]{Fryer1999,Heger2003}.

\section{Comparison to observations and preliminary predictions}
\label{sec:observations}

Estimates of the KE and $M_{\textrm{Ni}}$ for several observed supernovae are shown in Figure~\ref{fig:predictions}, where Type II (Type Ibc) supernovae, in which hydrogen is detected (not detected), are marked with black crosses (x-symbols). The observations, listed in Table~\ref{tbl:observations}, 
were compiled from \citet[][]{Hamuy2003,Hendry2005,Pastorello2005,Inserra2011,Pastorello2012,Taddia2012,Tomasella2013,Dall'Ora2014,Lyman2014,Spiro2014,UC2014}, and include only CCSNe within comoving radial distance of $100\,\textrm{Mpc}$ (to exclude rare events). The observations include modeling of the emitted light and are susceptible to systematic uncertainties. Moreover, the distribution of the sample in the KE--$M_{\textrm{Ni}}$ plane does not represent relative rates of events. Nevertheless, the range of observed KE ($10^{50}-\textrm{few}\times10^{52}\,\textrm{erg}$) and $M_{\textrm{Ni}}$ ($10^{-3}-1\,M_{\odot}$), and the gross correlation between them (higher KE leads to higher $M_{\textrm{Ni}}$), can be deduced. The lower limits of the KE and $M_{\textrm{Ni}}$ may be observationally biased. 

The calculated KE and $M_{\textrm{Ni}}$ from Section~\ref{sec:numerical} are compared to observations. Because of the sensitivity of the calculated results to the numerical treatment of the inner boundary (Section~\ref{sec:convergence}) and the partial parameter scan, this comparison, as well as the predictions made, are preliminary. As can be seen, the range of observed KE and $M_{\textrm{Ni}}$, as well as the gross correlation between them, can be obtained from CITE. It is also possible to obtain $\textrm{KE}\approx\textrm{few}\times10^{49}\,\textrm{erg}$ without tuning (see Section~\ref{sec:shell location}) for small values of $M_{\textrm{base}}$. One feature of CITE is that KE scales with $M_{\textrm{base}}$, i.e. stronger explosions originated from higher mass progenitors, as observations suggest \citep[][]{Poznanski2013}, and in contrast to the neutrino mechanism \cite[][]{Fryer1999,Heger2003}.

The calculated KE and remnant mass are shown in panel (b) of Figure~\ref{fig:predictions}. The remnant mass in the neutron star (NS) regime, $M_{\textrm{grav}}$, includes a correction due to the negative binding energy by subtracting an assumed gravitational energy of $1.5\cdot10^{53}(M_{\textrm{grav}}/M_{\odot})^{2}\,\textrm{erg}$ \citep[][]{LY1989}. For gravitational mass above the maximal mass of a NS (taken here as $2.5\,M_{\odot}$) the remnant mass represents the baryonic mass of the in-falling material. Stronger explosions are predicted to leave more massive remnants, in contrast to the neutrino mechanism \citep[where strong explosions leave neutron stars while weak explosions leave black holes;][]{Fryer1999,Heger2003}. CITE predicts that NSs can only be produced in weak ($\simlt10^{51}\,\textrm{erg}$) explosions \citep[explaining, e.g., the low, $<10^{50}\,\textrm{erg}$, KE of the Crab nebula;][]{Smith2013}, while strong ($\simgt10^{51}\,\textrm{erg}$) explosions must leave black holes (BHs). This indicates that the vast majority of observed extragalactic CCSNe (including SN1987A) left behind BHs.

\acknowledgments I thank B. Katz for a thorough reading of the manuscript and for useful discussions. I thank U. Nakar and D. Poznanski for useful discussions. D.~K. gratefully acknowledges support from the Friends of the Institute for Advanced Study. FLASH was in part developed by the DOE NNSA-ASC OASCR Flash Center at the University of Chicago. 


\bibliographystyle{apj}

\newpage

\begin{figure}
        \subfigure[]{
             \includegraphics[width=0.4\textwidth]{f4a.eps}
        }
        \subfigure[]{
             \includegraphics[width=0.4\textwidth]{f4b.eps}
        }
        \subfigure[]{
             \includegraphics[width=0.4\textwidth]{f4c.eps}
        }
        \subfigure[]{
             \includegraphics[width=0.4\textwidth]{f4d.eps}
        }
      \caption{Results of calculations with $t_{b,0}/t_{ff,0}=100$, $X_{\textrm{He}}=X_{\textrm{O}}=0.5$, $\Mshell=\Mshellmax$, $\frotz=0.05$, $\xrotz=0.8$, $M_{\textrm{base}}\in\left[1.5,16.5\right]M_{\odot}$, and $\rho_{\textrm{base}}\in\left[0.5,2.5\right]\cdot10^{4}\,\textrm{g}\,\textrm{cm}^{-3}$ (different colors represent different $\rho_{\textrm{base}}$). Panel (a): KE; panel (b): $M_{\textrm{Ni}}$; panel (c): in-falling mass; panel (d): KE as a function of $-E_{\textrm{bin}}$.}
   \label{fig:full}
\end{figure}

\begin{figure}
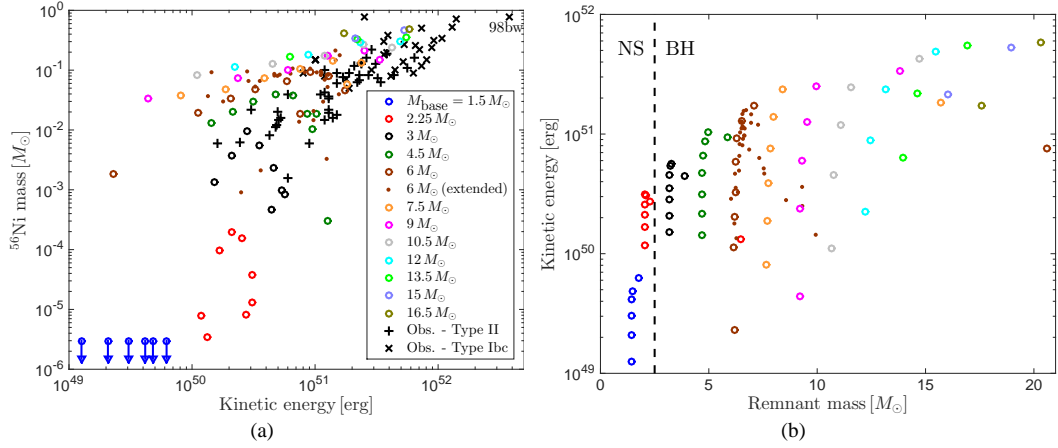

        \subfigure[]{
             \includegraphics[width=0.38\textwidth]{f5a.eps}
        }
        \subfigure[]{
             \includegraphics[width=0.38\textwidth]{f5b.eps}
        }
      \caption{The uniform sample of Figure~\ref{fig:full} (circles, different colors represent different $M_{\textrm{base}}$) extended with the sensitivity calculations of Figure~\ref{fig:example_param} (points). Panel (a): $M_{\textrm{Ni}}$--KE. The observed sample, black crosses (x-symbols) for type II (type Ibc), is shown without error bars (given in Table~\ref{tbl:observations}
) for purpose of clarity. As can be seen, the range of observed KE and $M_{\textrm{Ni}}$, as well as the gross correlation between them, can be obtained from CITE. Panel (b): KE--remnant mass (see Section~\ref{sec:observations} for the definition of the remnant mass).}
   \label{fig:predictions}
\end{figure}

\newpage

\begin{deluxetable}{cccccccccc}
\tablecaption{The set of simulations examining the sensitivity of the $M_{\textrm{base}}=6\,M_{\odot}$, $\rho_{\textrm{base}}=10^{4}\,\textrm{g}\,\textrm{cm}^{-3}$, $t_{b,0}/t_{ff,0}=100$, $X_{\textrm{He}}=0.5$, $\frotz=0.05$, $\xrotz=0.8$ calculation with $\Delta R=28.6\,\textrm{km}$, $r_{\textrm{res}}=5\cdot10^{9}\,\textrm{cm}$, $r_{\textrm{inner}}=5\cdot10^{7}\,\textrm{cm}$, to various numerical and physical parameters.\label{tbl:lista}}
\tablewidth{0pt}
\tablehead{ \colhead{description} & \colhead{$r_{\textrm{base}}$} & \colhead{$s_{\textrm{He-O}}$} & \colhead{$\Mshell$} & \colhead{$\textrm{total mass}$} & \colhead{$-E_{\textrm{bin}}$} & \colhead{$\textrm{kinetic energy}$} & \colhead{$^{56}\textrm{Ni mass}$} & \colhead{$\textrm{in-falling mass}$\tablenotemark{a}} \\
            \colhead{} &  \colhead{$[10^{9}\,\textrm{cm}]$} & \colhead{$[k_{B}]$} & \colhead{$[M_{\odot}]$} & \colhead{$[M_{\odot}]$} & \colhead{$[10^{51}\,\textrm{erg}]$} & \colhead{$[10^{51}\,\textrm{erg}]$} & \colhead{$[10^{-2}\,M_{\odot}]$} & \colhead{$[M_{\odot}]$}}
\startdata
reference	&	$4.62$	&	$7.81$	&	$3.60$	&	$11.0$	&	$0.856$	&	                $1.29$	&	$8.02$	&	$6.50$	\\
\hline
$\Delta R=14.3\,\textrm{km}$	&		&		&		&		&	                    &	$1.40$	&	$8.66$	&	$6.45$	\\
$\Delta R=57.2\,\textrm{km}$	&		&		&		&		&		                &	$1.37$	&	$5.18$	&	$6.61$	\\
$r_{\textrm{res}}=10^{10}\,\textrm{cm}$	&		&		&		&		&	           &	$1.33$	&	$8.33$	&	$6.46$	\\
$r_{\textrm{inner}}=100\,\textrm{km}$	&		&		&		&		&	           &	$1.54$	&	$16.2$	&	$6.31$	\\
$r_{\textrm{inner}}=250\,\textrm{km}$	&		&		&		&		&	           &	$1.40$	&	$11.1$	&	$6.38$	\\
$r_{\textrm{inner}}=1000\,\textrm{km}$	&		&		&		&		&	           &	$1.19$	&	$0.0147$&	$6.87$	\\
\hline
$\xrotz=0.4$	&		&		&		&		&	                                   &	$0.533$	&	$14.3$  &	$7.42$	\\
$\xrotz=0.5$	&		&		&		&		&	                                   &	$0.545$	&	$15.6$  &	$7.22$	\\
$\xrotz=0.6$	&		&		&		&		&	                                   &	$0.515$	&	$13.6$  &	$7.31$	\\
$\xrotz=0.7$	&		&		&		&		&	                                   &	$0.912$	&	$10.6$  &	$6.57$	\\
$\xrotz=0.9$	&		&		&		&		&	                                   &	$1.56$	&	$3.96$  &	$6.61$	\\
$\xrotz=1$   	&		&		&		&		&	                                   &	$1.24$	&	$0.329$ &	$6.95$	\\
$\xrotz=1.1$	&		&		&		&		&	                                   &	$0.25$	&	$0.0907$&	$9.31$	\\
\hline
$\frotz=0.015$	&		&		&		&		&	                                   &	$0.362$	&	$0.211$&	$9.30$	\\
$\frotz=0.02$	&		&		&		&		&	                                   &	$1.20$	&	$6.92$&	$7.27$	\\
$\frotz=0.025$	&		&		&		&		&	                                   &	$1.46$	&	$21.3$&	$6.95$	\\
$\frotz=0.03$	&		&		&		&		&	                                   &	$1.59$	&	$17.8$&	$6.72$	\\
$\frotz=0.04$	&		&		&		&		&	                                   &	$1.47$	&	$14.7$&	$6.55$	\\
$\frotz=0.06$	&		&		&		&		&	                                   &	$1.26$	&	$5.23$&	$6.52$	\\
$\frotz=0.07$	&		&		&		&		&	                                   &	$1.16$	&	$2.98$&	$6.55$	\\
$\frotz=0.08$	&		&		&		&		&	                                   &	$0.972$	&	$2.07$&	$6.63$	\\
$\frotz=0.09$	&		&		&		&		&	                                   &	$0.970$	&	$1.46$&	$6.65$	\\
$\frotz=0.1$	&		&		&		&		&	                                   &	$0.742$	&	$1.37$&	$6.86$	\\
\hline
$\textrm{relaxed}$	&		&		&		&		&	                                   &	$1.80$	&	$9.77$&	$6.73$	\\
$\textrm{relaxed},\,\frotz=0.1$	&		&		&		&		&	                                   &	$1.09$	&	$1.16$&	$6.79$	\\
\hline
$X_{\textrm{He}}=0.2$	&	$4.66$	& $7.00$ &	$3.84$ & $11.4$	&	$0.886$             &	$0.144$	&	$9.23$&	$9.93$	\\
$X_{\textrm{He}}=0.3$	&	$4.61$	& $7.23$ &	$3.56$ & $11.0$	&	$0.839$             &	$0.467$	&	$8.12$&	$7.56$	\\
$X_{\textrm{He}}=0.4$	&	$4.60$	& $7.51$ &	$3.52$ & $10.9$	&	$0.835$             &	$1.13$	&	$8.98$&	$6.61$	\\
$X_{\textrm{He}}=0.6$	&	$4.65$	& $8.13$ &	$3.78$ & $11.3$	&	$0.895$             &	$1.11$	&	$9.57$&	$6.65$	\\
$X_{\textrm{He}}=0.7$	&	$4.69$	& $8.46$ &	$4.06$ & $11.6$	&	$0.953$             &	$0.631$	&	$8.74$&	$7.33$	\\
$X_{\textrm{He}}=0.8$	&	$4.75$	& $8.83$ &	$4.50$ & $12.2$	&	$1.04$             &	$0.288$	&	$6.59$&	$6.02$	\\
\hline
                    	&	$2.71$	&	$8.19$	&	$0.444$	&	$6.72$	&	$0.173$	&	   $0.135$	&	$3.68$	&	$6.27$	\\
                    	&	$3.23$	&	$8.06$	&	$0.929$	&	$7.47$	&	$0.295$	&	   $0.326$	&	$5.32$	&	$6.34$	\\
                    	&	$3.58$	&	$7.99$	&	$1.42$	&	$8.19$	&	$0.404$	&	   $0.495$	&	$8.03$	&	$6.39$	\\
                    	&	$3.86$	&	$7.93$	&	$1.89$	&	$8.86$	&	$0.506$	&	   $0.657$	&	$10.7$	&	$6.41$	\\
                    	&	$4.09$	&	$7.89$	&	$2.36$	&	$9.48$	&	$0.602$	&	   $0.824$	&	$9.20$	&	$6.42$	\\
                    	&	$4.30$	&	$7.86$	&	$2.81$	&	$10.1$	&	$0.694$	&	   $1.03$	&	$9.18$	&	$6.44$	\\                                        	 & 	    $4.48$	&	$7.83$	&	$3.25$	&	$10.6$	&	$0.783$	&	   $1.11$	&	$8.28$	&	$6.51$	\\
\hline
$t_{b,0}/t_{ff,0}=10^{3}$	&	$3.01$	& $7.18$ &	$0.389$ & $6.55$	&	$0.157$             &	$0.179$	&	$2.99$ &	$6.18$	\\
$t_{b,0}/t_{ff,0}=10^{3}$	&	$3.59$	& $7.08$ &	$0.799$ & $7.11$	&	$0.267$             &	$0.355$	&	$6.59$ &	$6.24$	\\
$t_{b,0}/t_{ff,0}=10^{3}$	&	$3.82$	& $7.05$ &	$1.02$ & $7.40$	&	$0.320$             &	$0.466$	&	$8.66$ &	$6.21$	\\
\enddata
\tablenotetext{a}{Without the binding energy correction used in panel (b) of Figure~\ref{fig:predictions}.}
\end{deluxetable}

\newpage

\begin{deluxetable}{cccccccccc}
\tablecaption{The set of simulations with $t_{b,0}/t_{ff,0}=100$, $X_{\textrm{He}}=0.5$, $\Mshell=\Mshellmax$, $\frotz=0.05$, and $\xrotz=0.8$.\label{tbl:listb}}
\tablewidth{0pt}
\tablehead{ \colhead{$M_{\textrm{base}}$} &  \colhead{$\rho_{\textrm{base}}$} & \colhead{$r_{\textrm{base}}$} & \colhead{$s_{\textrm{He-O}}$} & \colhead{$\Mshell$} & \colhead{$\textrm{total mass}$} & \colhead{$-E_{\textrm{bin}}$} & \colhead{$\textrm{kinetic energy}$} & \colhead{$^{56}\textrm{Ni mass}$} & \colhead{$\textrm{in-falling mass}$\tablenotemark{a}} \\
            \colhead{$[M_{\odot}]$} &  \colhead{$[10^{4}\,\textrm{g}\,\textrm{cm}^{-3}]$} & \colhead{$[10^{9}\,\textrm{cm}]$} & \colhead{$[k_{B}]$} & \colhead{$[M_{\odot}]$} & \colhead{$[M_{\odot}]$} & \colhead{$[10^{51}\,\textrm{erg}]$} & \colhead{$[10^{51}\,\textrm{erg}]$} & \colhead{$[10^{-2}\,M_{\odot}]$} & \colhead{$[M_{\odot}]$}}
\startdata
$1.5$	&	$1.5$	&	$1.50$	&	$7.06$	&	$0.540$	&	$2.97$	&	$0.0714$	&	$0.0621$	&	-	&	$2.01$	\\
$1.5$	&	$1.75$	&	$1.36$	&	$6.75$	&	$0.297$	&	$2.18$	&	$0.0443$	&	$0.0481$	&	-	&	$1.69$	\\
$1.5$	&	$2$	&	$1.25$	&	$6.51$	&	$0.187$	&	$1.85$	&	$0.0319$	&	$0.0416$	&	-	&	$1.64$	\\
$1.5$	&	$2.25$	&	$1.16$	&	$6.31$	&	$0.130$	&	$1.71$	&	$0.0252$	&	$0.0305$	&	-	&	$1.63$	\\
$1.5$	&	$2.5$	&	$1.09$	&	$6.15$	&	$0.0964$	&	$1.64$	&	$0.0210$	&	$0.0207$	&	-	&	$1.63$	\\
$1.5$	&	$2.75$	&	$1.03$	&	$6.01$	&	$0.0755$	&	$1.60$	&	$0.0182$	&	$0.0126$	&	-	&	$1.63$	\\
\hline
$2.25$	&	$1$	&	$3.00$	&	$7.88$	&	$3.91$	&	$9.06$	&	$0.459$	&	$0.133$	&	$0.000347$	&	$6.48$	\\
$2.25$	&	$1.25$	&	$2.63$	&	$7.31$	&	$2.21$	&	$6.07$	&	$0.282$	&	$0.274$	&	$0.000804$	&	$2.72$	\\
$2.25$	&	$1.5$	&	$2.35$	&	$6.91$	&	$1.33$	&	$4.43$	&	$0.188$	&	$0.305$	&	$0.00130$	&	$2.46$	\\
$2.25$	&	$1.75$	&	$2.12$	&	$6.62$	&	$0.842$	&	$3.54$	&	$0.135$	&	$0.311$	&	$0.00378$	&	$2.42$	\\
$2.25$	&	$2$	&	$1.94$	&	$6.39$	&	$0.567$	&	$3.05$	&	$0.103$	&	$0.255$	&	$0.0153$	&	$2.41$	\\
$2.25$	&	$2.25$	&	$1.79$	&	$6.20$	&	$0.403$	&	$2.78$	&	$0.0829$	&	$0.211$	&	$0.0198$	&	$2.40$	\\
$2.25$	&	$2.5$	&	$1.67$	&	$6.05$	&	$0.300$	&	$2.62$	&	$0.0689$	&	$0.167$	&	$0.00950$	&	$2.40$	\\
$2.25$	&	$2.75$	&	$1.56$	&	$5.92$	&	$0.232$	&	$2.53$	&	$0.0588$	&	$0.118$	&	$0.000793$	&	$2.40$	\\
\hline
$3$	&	$1$	&	$3.57$	&	$7.83$	&	$4.11$	&	$9.52$	&	$0.575$	&	$0.444$	&	$0.0462$	&	$3.90$	\\
$3$	&	$1.25$	&	$3.12$	&	$7.27$	&	$2.38$	&	$6.64$	&	$0.370$	&	$0.568$	&	$0.0839$	&	$3.27$	\\
$3$	&	$1.5$	&	$2.76$	&	$6.88$	&	$1.45$	&	$5.10$	&	$0.256$	&	$0.537$	&	$0.0981$	&	$3.21$	\\
$3$	&	$1.75$	&	$2.48$	&	$6.59$	&	$0.931$	&	$4.26$	&	$0.187$	&	$0.459$	&	$0.229$	&	$3.19$	\\
$3$	&	$2$	&	$2.25$	&	$6.37$	&	$0.627$	&	$3.80$	&	$0.144$	&	$0.350$	&	$0.550$	&	$3.18$	\\
$3$	&	$2.25$	&	$2.06$	&	$6.19$	&	$0.442$	&	$3.54$	&	$0.115$	&	$0.281$	&	$0.956$	&	$3.16$	\\
$3$	&	$2.5$	&	$1.90$	&	$6.04$	&	$0.324$	&	$3.38$	&	$0.0945$	&	$0.209$	&	$0.377$	&	$3.16$	\\
$3$	&	$2.75$	&	$1.76$	&	$5.91$	&	$0.246$	&	$3.28$	&	$0.0797$	&	$0.152$	&	$0.135$	&	$3.16$	\\
\hline
$4.5$	&	$0.75$	&	$5.03$	&	$8.72$	&	$7.54$	&	$16.0$	&	$1.26$	&	$0.942$	&	$1.04$	&	$5.86$	\\
$4.5$	&	$1$	&	$4.24$	&	$7.81$	&	$3.98$	&	$10.3$	&	$0.746$	&	$1.04$	&	$1.86$	&	$4.97$	\\
$4.5$	&	$1.25$	&	$3.64$	&	$7.25$	&	$2.22$	&	$7.56$	&	$0.480$	&	$0.872$	&	$1.87$	&	$4.81$	\\
$4.5$	&	$1.5$	&	$3.16$	&	$6.88$	&	$1.28$	&	$6.17$	&	$0.327$	&	$0.665$	&	$3.75$	&	$4.75$	\\
$4.5$	&	$1.75$	&	$2.78$	&	$6.60$	&	$0.773$	&	$5.45$	&	$0.232$	&	$0.477$	&	$3.85$	&	$4.71$	\\
$4.5$	&	$2$	&	$2.46$	&	$6.39$	&	$0.487$	&	$5.07$	&	$0.171$	&	$0.312$	&	$2.95$	&	$4.70$	\\
$4.5$	&	$2.25$	&	$2.2$	&	$6.21$	&	$0.321$	&	$4.87$	&	$0.130$	&	$0.215$	&	$1.98$	&	$4.68$	\\
$4.5$	&	$2.5$	&	$1.99$	&	$6.07$	&	$0.222$	&	$4.75$	&	$0.103$	&	$0.143$	&	$1.33$	&	$4.67$	\\
\hline
$6$	&	$0.5$	&	$6.91$	&	$10.6$	&	$15.9$	&	$30.8$	&	$2.95$	&	$0.763$	&	$1.86$	&	$20.6$	\\
$6$	&	$0.75$	&	$5.57$	&	$8.71$	&	$7.24$	&	$16.7$	&	$1.48$	&	$1.74$	&	$4.56$	&	$7.10$	\\
$6$	&	$1$	&	$4.62$	&	$7.81$	&	$3.60$	&	$11.0$	&	$0.856$	&	$1.29$	&	$8.02$	&	$6.50$	\\
$6$	&	$1.25$	&	$3.88$	&	$7.27$	&	$1.85$	&	$8.43$	&	$0.528$	&	$0.917$	&	$9.13$	&	$6.30$	\\
$6$	&	$1.5$	&	$3.28$	&	$6.91$	&	$0.965$	&	$7.20$	&	$0.338$	&	$0.588$	&	$6.44$	&	$6.26$	\\
$6$	&	$1.75$	&	$2.79$	&	$6.64$	&	$0.519$	&	$6.61$	&	$0.222$	&	$0.328$	&	$4.68$	&	$6.22$	\\
$6$	&	$2$	&	$2.40$	&	$6.43$	&	$0.294$	&	$6.33$	&	$0.152$	&	$0.205$	&	$3.41$	&	$6.18$	\\
$6$	&	$2.25$	&	$2.10$	&	$6.27$	&	$0.178$	&	$6.20$	&	$0.109$	&	$0.113$	&	$1.94$	&	$6.16$	\\
$6$	&	$2.5$	&	$1.86$	&	$6.12$	&	$0.115$	&	$6.13$	&	$0.0817$	&	$0.0231$	&	$0.185$	&	$6.18$	\\
\hline
$7.5$	&	$0.5$	&	$7.47$	&	$10.5$	&	$15.7$	&	$31.7$	&	$3.34$	&	$1.82$	&	$5.75$	&	$15.7$	\\
$7.5$	&	$0.75$	&	$5.94$	&	$8.72$	&	$6.82$	&	$17.3$	&	$1.65$	&	$2.36$	&	$13.0$	&	$8.40$	\\
$7.5$	&	$1$	&	$4.83$	&	$7.83$	&	$3.14$	&	$11.8$	&	$0.915$	&	$1.40$	&	$14.4$	&	$8.00$	\\
$7.5$	&	$1.25$	&	$3.94$	&	$7.30$	&	$1.43$	&	$9.32$	&	$0.525$	&	$0.759$	&	$10.6$	&	$7.83$	\\
$7.5$	&	$1.5$	&	$3.20$	&	$6.95$	&	$0.642$	&	$8.27$	&	$0.303$	&	$0.388$	&	$7.43$	&	$7.76$	\\
$7.5$	&	$1.75$	&	$2.62$	&	$6.70$	&	$0.298$	&	$7.84$	&	$0.180$	&	$0.188$	&	$4.77$	&	$7.70$	\\
$7.5$	&	$2$	&	$2.19$	&	$6.50$	&	$0.151$	&	$7.67$	&	$0.114$	&	$0.0813$	&	$3.77$	&	$7.67$	\\
\hline
$9$	&	$0.5$	&	$7.92$	&	$10.5$	&	$15.4$	&	$32.4$	&	$3.70$	&	$3.36$	&	$15.1$	&	$13.8$	\\
$9$	&	$0.75$	&	$6.21$	&	$8.74$	&	$6.33$	&	$18.0$	&	$1.77$	&	$2.52$	&	$21.2$	&	$9.94$	\\
$9$	&	$1$	&	$4.92$	&	$7.86$	&	$2.64$	&	$12.5$	&	$0.928$	&	$1.26$	&	$17.6$	&	$9.55$	\\
$9$	&	$1.25$	&	$3.85$	&	$7.35$	&	$1.02$	&	$10.3$	&	$0.478$	&	$0.602$	&	$9.91$	&	$9.31$	\\
$9$	&	$1.5$	&	$2.98$	&	$7.02$	&	$0.377$	&	$9.45$	&	$0.241$	&	$0.238$	&	$7.47$	&	$9.23$	\\
$9$	&	$1.75$	&	$2.35$	&	$6.77$	&	$0.153$	&	$9.17$	&	$0.130$	&	$0.0441$	&	$3.30$	&	$9.19$	\\
\hline
$10.5$	&	$0.5$	&	$8.29$	&	$10.6$	&	$15.1$	&	$33.2$	&	$4.02$	&	$4.24$	&	$23.6$	&	$14.7$	\\
$10.5$	&	$0.75$	&	$6.39$	&	$8.75$	&	$5.80$	&	$18.6$	&	$1.86$	&	$2.48$	&	$26.6$	&	$11.6$	\\
$10.5$	&	$1$	&	$4.91$	&	$7.90$	&	$2.13$	&	$13.3$	&	$0.897$	&	$1.18$	&	$17.3$	&	$11.1$	\\
$10.5$	&	$1.25$	&	$3.64$	&	$7.41$	&	$0.661$	&	$11.3$	&	$0.398$	&	$0.451$	&	$12.6$	&	$10.8$	\\
$10.5$	&	$1.5$	&	$2.67$	&	$7.10$	&	$0.200$	&	$10.7$	&	$0.175$	&	$0.110$	&	$8.15$	&	$10.7$	\\
\hline
$12$	&	$0.5$	&	$8.6$	&	$10.6$	&	$14.7$	&	$33.9$	&	$4.31$	&	$4.93$	&	$29.8$	&	$15.5$	\\
$12$	&	$0.75$	&	$6.51$	&	$8.78$	&	$5.24$	&	$19.3$	&	$1.91$	&	$2.35$	&	$28.9$	&	$13.2$	\\
$12$	&	$1$	&	$4.8$	&	$7.94$	&	$1.63$	&	$14.1$	&	$0.825$	&	$0.880$	&	$18.1$	&	$12.5$	\\
$12$	&	$1.25$	&	$3.32$	&	$7.49$	&	$0.388$	&	$12.5$	&	$0.303$	&	$0.225$	&	$11.4$	&	$12.2$	\\
\hline
$13.5$	&	$0.5$	&	$8.86$	&	$10.6$	&	$14.2$	&	$34.7$	&	$4.57$	&	$5.48$	&	$35.4$	&	$16.9$	\\
$13.5$	&	$0.75$	&	$6.56$	&	$8.81$	&	$4.67$	&	$19.9$	&	$1.92$	&	$2.19$	&	$32.8$	&	$14.6$	\\
$13.5$	&	$1$	&	$4.59$	&	$8.00$	&	$1.17$	&	$15.0$	&	$0.716$	&	$0.630$	&	$16.7$	&	$14.0$	\\
\hline
$15$	&	$0.5$	&	$9.08$	&	$10.6$	&	$13.8$	&	$35.4$	&	$4.79$	&	$5.34$	&	$46.3$	&	$19.0$	\\
$15$	&	$0.75$	&	$6.57$	&	$8.84$	&	$4.10$	&	$20.6$	&	$1.89$	&	$2.13$	&	$33.7$	&	$16.0$	\\
\hline
$16.5$	&	$0.5$	&	$9.27$	&	$10.6$	&	$13.3$	&	$36.1$	&	$4.99$	&	$5.81$	&	$47.6$	&	$20.3$	\\
$16.5$	&	$0.75$	&	$6.51$	&	$8.87$	&	$3.53$	&	$21.3$	&	$1.83$	&	$1.71$	&	$40.9$	&	$17.6$	\\
\enddata
\tablenotetext{a}{Without the binding energy correction used in panel (b) of Figure~\ref{fig:predictions}.}
\end{deluxetable}

\newpage

\begin{deluxetable}{cccccccccc}
\tablecaption{A compilation from the literature of estimated KE and $M_{\textrm{Ni}}$ from the light curves\label{tbl:observations}}
\tablewidth{0pt}
\tablehead{ \colhead{Name} &  \colhead{$\textrm{Kinetic energy}\,[10^{51}\,\textrm{erg}]$} & \colhead{$^{56}\textrm{Ni mass}\,[M_{\odot}]$} & \colhead{Type} & \colhead{Reference} & \colhead{Name} &  \colhead{$\textrm{Kinetic energy}\,[10^{51}\,\textrm{erg}]$} & \colhead{$^{56}\textrm{Ni mass}\,[M_{\odot}]$} & \colhead{Type} & \colhead{Reference}}
\startdata

69L	&	$2.3^{+0.7}_{-0.6}$	&	$0.082^{+0.034}_{-0.026}$	&	II	&	3	&	73R	&	$2.7^{+1.2}_{-0.9}$	&	$0.084^{+0.044}_{-0.03}$	&	II	&	3	\\
83I	&	$1$	&	$0.15$	&	Ibc	&	3	&	83N	&	$1$	&	$0.15$	&	Ibc	&	3	\\
84L	&	$1$	&	$0.15$	&	Ibc	&	3	&	86L	&	$1.3^{+0.5}_{-0.3}$	&	$0.034^{+0.018}_{-0.011}$	&	II	&	3	\\
87A	&	$1.7$	&	$0.075$	&	II	&	3	&	88A	&	$2.2^{+1.7}_{-1.2}$	&	$0.062^{+0.029}_{-0.02}$	&	II	&	3	\\
89L	&	$1.2^{+0.6}_{-0.5}$	&	$0.015^{+0.008}_{-0.005}$	&	II	&	3	&	90E	&	$3.4^{+1.3}_{-1}$	&	$0.062^{+0.031}_{-0.022}$	&	II	&	3	\\
91G	&	$1.3^{+0.9}_{-0.6}$	&	$0.022^{+0.008}_{-0.006}$	&	II	&	3	&	92H	&	$3.1^{+1.3}_{-1}$	&	$0.129^{+0.053}_{-0.037}$	&	II	&	3	\\
92ba	&	$1.3^{+0.5}_{-0.4}$	&	$0.019^{+0.009}_{-0.007}$	&	II	&	3	&	93J	&	$2.4^{+1.1}_{-1}$	&	$0.13^{+0.02}_{-0.01}$	&	II	&	11	\\
94I	&	$1.2^{+0.6}_{-0.5}$	&	$0.08^{+0.01}_{-0.01}$	&	Ibc	&	11	&	96cb	&	$2.1^{+1.6}_{-0.9}$	&	$0.12^{+0.04}_{-0.03}$	&	II	&	11	\\
97D	&	$0.9$	&	$0.006$	&	II	&	3	&	97ef	&	$8$	&	$0.15$	&	Ibc	&	3	\\
98A	&	$5.6$	&	$0.11$	&	II	&	7	&	98bw	&	$38.2^{+13}_{-11.1}$	&	$0.76^{+0.11}_{-0.1}$	&	Ibc	&	11	\\
99br	&	$0.6$	&	$0.0016^{+0.0011}_{-0.0008}$	&	II	&	3	&	99cr	&	$1.9^{+0.8}_{-0.6}$	&	$0.09^{+0.034}_{-0.027}$	&	II	&	3	\\
99dn	&	$7.3^{+2.6}_{-3.6}$	&	$0.12^{+0.01}_{-0.02}$	&	Ibc	&	11	&	99em	&	$1.3^{+0.1}_{-0.1}$	&	$0.036^{+0.009}_{-0.009}$	&	II	&	1	\\
99em	&	$1.2^{+0.6}_{-0.3}$	&	$0.042^{+0.027}_{-0.019}$	&	II	&	3	&	99ex	&	$3.6^{+2.1}_{-1.5}$	&	$0.18^{+0.05}_{-0.04}$	&	Ibc	&	11	\\
99gi	&	$1.5^{+0.7}_{-0.5}$	&	$0.018^{+0.013}_{-0.009}$	&	II	&	3	&	00cb	&	$4.4^{+0.3}_{-0.3}$	&	$0.083^{+0.039}_{-0.039}$	&	II	&	1	\\
02ap	&	$6.3^{+3.8}_{-2.9}$	&	$0.09^{+0.01}_{-0.01}$	&	Ibc	&	11	&	03Z	&	$0.245^{+0.018}_{-0.018}$	&	$0.0063^{+0.0006}_{-0.0006}$	&	II	&	4	\\
03bg	&	$3.8^{+1.8}_{-1.6}$	&	$0.19^{+0.03}_{-0.02}$	&	II	&	11	&	03gd	&	$1.4^{+0.3}_{-0.3}$	&	$0.016^{+0.01}_{-0.006}$	&	II	&	5	\\
03jd	&	$7.4^{+2.8}_{-2.4}$	&	$0.51^{+0.1}_{-0.09}$	&	Ibc	&	11	&	04aw	&	$6.6^{+2.3}_{-3.3}$	&	$0.26^{+0.05}_{-0.04}$	&	Ibc	&	11	\\
04dk	&	$5.3^{+3}_{-2.2}$	&	$0.27^{+0.05}_{-0.04}$	&	Ibc	&	11	&	04dn	&	$7.1^{+3.5}_{-3.6}$	&	$0.22^{+0.04}_{-0.03}$	&	Ibc	&	11	\\
04et	&	$2.3^{+0.3}_{-0.3}$	&	$0.068^{+0.009}_{-0.009}$	&	II	&	1	&	04fe	&	$3.6^{+1.5}_{-1.7}$	&	$0.3^{+0.05}_{-0.05}$	&	Ibc	&	11	\\
04ff	&	$2.9^{+1.6}_{-1.9}$	&	$0.22^{+0.04}_{-0.03}$	&	II	&	11	&	04gq	&	$5.2^{+2.9}_{-2.2}$	&	$0.14^{+0.07}_{-0.05}$	&	Ibc	&	11	\\
05az	&	$3.9^{+2.5}_{-1.7}$	&	$0.38^{+0.08}_{-0.07}$	&	Ibc	&	11	&	05bf	&	$0.8^{+1.4}_{-0.3}$	&	$0.09^{+0.04}_{-0.02}$	&	Ibc	&	11	\\
05cs	&	$0.43^{+0.03}_{-0.03}$	&	$0.0082^{+0.0016}_{-0.0016}$	&	II	&	1	&	05cs	&	$0.16^{+0.03}_{-0.03}$	&	$0.006^{+0.003}_{-0.003}$	&	II	&	2	\\
05hg	&	$2.5^{+1.1}_{-1.2}$	&	$0.76^{+0.11}_{-0.1}$	&	Ibc	&	11	&	06T	&	$1.2^{+0.6}_{-0.5}$	&	$0.1^{+0.04}_{-0.02}$	&	II	&	11	\\
06au	&	$3.2$	&	$0.073$	&	II	&	8	&	06el	&	$6.4^{+2.6}_{-4.1}$	&	$0.16^{+0.03}_{-0.03}$	&	II	&	11	\\
06ep	&	$4.1^{+2.2}_{-2.4}$	&	$0.08^{+0.03}_{-0.02}$	&	Ibc	&	11	&	06ov	&	$2.4$	&	$0.127$	&	II	&	8	\\
07C	&	$3.8^{+1.6}_{-2.3}$	&	$0.2^{+0.05}_{-0.04}$	&	Ibc	&	11	&	07Y	&	$1.9^{+1.8}_{-1}$	&	$0.05^{+0.01}_{-0.01}$	&	Ibc	&	11	\\
07gr	&	$2.9^{+1.3}_{-1.1}$	&	$0.1^{+0.02}_{-0.01}$	&	Ibc	&	11	&	07od	&	$0.5$	&	$0.02$	&	II	&	6	\\
07ru	&	$13^{+6.2}_{-7.3}$	&	$0.52^{+0.05}_{-0.05}$	&	Ibc	&	11	&	07uy	&	$10.8^{+3.7}_{-5.9}$	&	$0.34^{+0.05}_{-0.04}$	&	Ibc	&	11	\\
08D	&	$4.5^{+3.7}_{-1.7}$	&	$0.1^{+0.02}_{-0.01}$	&	Ibc	&	11	&	08ax	&	$2.6^{+2.9}_{-1.1}$	&	$0.16^{+0.05}_{-0.04}$	&	II	&	11	\\
08in	&	$0.505^{+0.34}_{-0.34}$	&	$0.015^{+0.005}_{-0.005}$	&	II	&	1	&	08in	&	$0.49^{+0.098}_{-0.098}$	&	$0.012^{+0.005}_{-0.005}$	&	II	&	2	\\
09E	&	$0.6$	&	$0.04$	&	II	&	7	&	09bb	&	$9.2^{+6}_{-3.2}$	&	$0.31^{+0.05}_{-0.04}$	&	Ibc	&	11	\\
09bw	&	$0.3$	&	$0.022$	&	II	&	10	&	09jf	&	$8.9^{+7.5}_{-4.3}$	&	$0.24^{+0.03}_{-0.02}$	&	Ibc	&	11	\\
11bm	&	$14^{+5.7}_{-5.6}$	&	$0.71^{+0.11}_{-0.09}$	&	Ibc	&	11	&	11dh	&	$1.5^{+0.8}_{-0.7}$	&	$0.09^{+0.01}_{-0.01}$	&	II	&	11	\\
11hs	&	$1.1^{+1}_{-0.5}$	&	$0.04^{+0.01}_{-0.01}$	&	II	&	11	&	12A	&	$0.48$	&	$0.011$	&	II	&	9	\\
12A	&	$0.525^{+0.06}_{-0.06}$	&	$0.016^{+0.002}_{-0.002}$	&	II	&	1	&	12aw	&	$1.5$	&	$0.06$	&	II	&	10	\\
iPTF13bvn	&	$1.8^{+0.8}_{-0.8}$	&	$0.07^{+0.02}_{-0.02}$	&	Ibc	&	11	&			
\enddata
\tablecomments{REFERENCES.--(1) \citet[][]{UC2014};(2) \citet[][]{Spiro2014};(3) \citet[][]{Hamuy2003};(4) \citet[][]{Hendry2005};(5) \citet[][]{Inserra2011};(6) \citet[][]{Pastorello2012};(7) \citet[][]{Pastorello2005};(8) \citet[][]{Taddia2012};(9) \citet[][]{Tomasella2013};(10) \citet[][]{Dall'Ora2014};(11) \citet[][]{Lyman2014}}
\end{deluxetable}

\end{document}